# ELASTIC COMPOSITE REINFORCED LIGHTWEIGHT CONCRETE AS A TYPE OF RESILIENT COMPOSITE SYSTEMS

ESMAEILI KAMYAR
*Research & Development Department of Nogamsazegan, Iran.*
newstructure1@gmail.com

*ABSTRACT—* A kind of Elastic Composite, Reinforced Lightweight Concrete (ECRLC) with the mentioned specifics is a type of "Resilient Composite Systems (RCS)" in which, contrary to the basic geometrical assumption of flexure theory in Solid Mechanics, "the strain changes in the beam height during bending" is typically "Non-linear".

Through employing this integrated structure, with significant high strain capability and modulus of resilience in bending, we could constructively achieve high bearing capacities in beams with secure fracture pattern, in less weight.

Due to the system's particulars and its behavior in bending, the usual calculation of the equilibrium steel amount to attain the low-steel bending sections with secure fracture pattern in the beams and its related limitations do not become propounded. Thereby, the strategic deadlock of high possibility of brittle fracture pattern in the bending elements made of the usual reinforced lightweight concretes, especially about the low-thickness bending elements as slabs, is being unlocked.

This simple, applied technology and the related components and systems can have several applications in "the Road and Building Industries" too.

Regarding the "strategic importance of the Lightweight & Integrated Construction in practical increase of the resistance and safety against earthquake" and considering the appropriate behavior of this resilient structure against the dynamic loads, shakes, impacts and shocks and capability of making some lightweight and insulating, non-brittle, reinforced sandwich panels and pieces, this system and its components could be also useful in "seismic areas".

This system could be also employed in constructing the vibration and impact absorber bearing pieces and slabs, which can be used in "the Railroad & Subway Structures" too.

Here, the "Resilient Composite Systems (RCS)" and particularly, ECRLC as a type of RCS have been concisely presented. [Meanwhile, in the related pictures & figures, an instance of the said new structure and its components and the results of some performed experiments (as the "in-bending" & in-compressive loadings of the slabs including this structure, similar to ASTM E 72 Standard) have been pointed.]

Keywords: Strength of materials (solid mechanics), Civil (construction), Materials, Earthquake (resistance and safety), Resilient concrete (flexible concrete, bendable concrete, elastic concrete), Composite concrete, Lightweight concrete, Reinforced concrete, Fibered concrete, Lightweight and integrated construction, Rail (railroad, railway), Subway, Road, Bridge, Resilience, Energy absorption, Fracture pattern, Non-linear, Strain changes, Beam, Ductility, Toughness, Insulating (insulation), Thin, Slab, Roof, Ceiling, Wall (partition), Building, Tower, Plan of mixture, Insulating reinforced lightweight pieces, 3d, Sandwich panel, Dry mix, Plaster, Foam, Expanded polystyrene (eps), Polypropylene, Pozzolan, Porous matrix (Pored matrix), Mesh (lattice), Cement, RCS, ECRLC

## I. INTRODUCTION

It is clear that; there are several advantages in employing lightweight concretes. In spite of these important advantages, there are numerous common and sometimes strategic problems and restrictions in using the lightweight concretes (especially about the lightweight concretes with oven-dry densities of $\leq$ 800kg/m$^3$ as the insulating concretes). These problems are among them: the shape of stress-strain diagram in the usual reinforced lightweight concretes and high possibility of brittle and non-secure being of fracture pattern; low mechanical strengths as the compressive, bending, tensile, and shearing strengths (e.g., punch shear); low ratios of "the dynamic and static elasticity modulus and shearing and tensile strengths" to "the compressive strength"; reinforcements inappropriate involvement in the usual lightweight concretes; volume instability and high shrinkage and contraction amounts and the problems resulted from loss, creep and fatigue; difficulties related to the lightweight concrete and reinforcements' durability (particularly in some environmental conditions in long-term); the problems related to the lateral forces transferring; some in-place implementation limitations and administrative restrictions; etc. [1], [2]

In planning the mentioned simple, applied technology and considering the option of appropriately employing some supplementary elements with the said composite system, here, it is attempted to concomitantly solving some of the said problems in the framework of "an integrated functioning unit" with "significant modulus of resilience (energy absorption capacity) and resistivity (specific strength as the ratio of the strength to the density) in bending", "non-brittle fracture pattern", and appropriate cost price.

## II. WHAT ARE THE RESILENT COMPOSITE SYSTEMS

A kind of "Elastic Composite, Reinforced Lightweight Concrete (ECRLC)" with the mentioned specifics, is a fibro-elastic, reinforced lightweight concrete having reticular structure. Indeed, this structure is a type of particular composite (compound) systems generally called as; "Resilient Composite Systems; R.C.S.". In the said composite systems, contrary to the basic geometrical assumption of flexure theory in the Solid Mechanics, the strain changes in the beam height during bending [3], is typically "Non-linear".

*A. General View*

As it was pointed; the "Resilient Compound Systems" are the complex materials with particular structural properties, in which, contrary to the basic geometrical assumption of flexure theory in the Solid Mechanics, the strain changes in the beam height during bending is typically "Non-linear".

Generally, the "Resilient Composite Systems (RCS)" are made by creating disseminated suitable hollow pores and/or by distributing appropriate lightweight aggregates in the supported reinforced, fibered conjoined matrix *so that* "the strain changes in the beam height during bending" is typically "non-linear", which has its own criteria and indices. Indeed, this is a particular method for making the compound materials or systems also named as "Resilient Composite Systems" having typically non-linear strain changes in the beam height during bending so that it leads to "less possibility of beam fracture of primary compressive type" and "more modulus of resilience" in bending, in "less weight (density)", in the said compound materials "with their own structural properties and specific functional criteria". [Here, the general term of "lightweight aggregate" has a broad meaning, also including the polymeric and non-polymeric beads or particles.]

In the "Resilient Composite Systems" in general, the main strategy to raise the modulus of resilience in bending is "increasing the strain capability of the system in bending" within the elastic limit.

Here, the main method or axial tactic to fulfill the stated strategy includes "creating suitable hollow pores and/or using appropriate lightweight aggregates, all disseminated in the matrix", for providing more possibility of expedient internal shape changes (deformities) in the matrix, which could lead to more appropriate distribution of the stresses and strains throughout the system. Conversely, only creating hollow pores and/or using the lightweight aggregates in the matrix, "by itself", not only won't lead to the mentioned goals, but also will bring about weakness and fragility of the matrix! Hence, concomitantly, the matrix should be supported and strengthened. Here, essentially strengthening and ameliorating are performed by giving attention to the internal consistency of the matrix and also through employing the reinforcements in "two complementary levels": 1- Using the fibers for better distribution of the tensile stresses and strains in the matrix and increase of the matrix's endurance and modulus of resilience in tension and bending; 2- Using the mesh or lattice for better distribution of the tensile stresses and strains in the system and increase of the system's endurance and modulus of resilience in tension and bending.

In these systems, the presence of the mentioned disseminated hollow pores and/or lightweight aggregates in the conjoined matrix (which has been ameliorated through forming an integrated, reticular structure) provides the possibility of "more internal deformities in the matrix" during bending. By the way, this could lead to less accumulation of the internal stresses in the certain points of the matrix during bending, better absorption and control of the stresses, and providing the possibility of more continuing the bending course particularly within the elastic limit.

Occurrence of the stated internal deformities in the system's supported matrix during bending also includes occurrence of the deformities in the mentioned hollow pores and/or lightweight aggregates disseminated in the conjoined matrix in two different forms. Indeed, we have the internal deformities in the fibered lightweight matrix of the system throughout the bending course in two main different forms: 1- Tendency to increase of the in-compressing layers thickness (height) (particularly in the upper parts of the beam) and conversion of some internal compressive stresses to the internal tensile stresses (in the axis perpendicular to the mentioned internal compressive tensions) in the in-compressing layers; 2- Tendency to decrease of the in-tension layers thickness (height) (particularly in the lower parts of the beam), and conversion of some internal tensile stresses to the internal compressive stresses (in the axis perpendicular to the mentioned internal tensile tensions) in the in-tension layers.

In the under-bending sections of the "Resilient Composite Systems", the established deformities in the "conjoined and perpendicular to load applying direction layers" during bending are so that "the initially plane and perpendicular to beam axis sections" typically remove from "the plane and vertical state" to "the curve shape" during bending ( 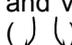 ). Thereby, the basic geometrical assumption of flexure theory in the Solid Mechanics ("linear" being of the strain changes in the beam height during bending) and its resulted trigonometric equations & equalities [3], [4] are being overshadowed in these systems.

In this way, through occurring of the stated internal deformities in the strengthened matrix, the stresses are more "distributed" and "absorbed" and the "rate" of increasing of the internal stresses in the matrix (could lead to the plasticity and crush of the matrix) are reduced. Indeed, in these systems, the mentioned internal deformities bring about the tendency of the so-called Neutral Axis to move downward. "This tendency is opposite to the natural tendency of the neutral axis to move upward during bending." Hence, more possibility for continuing the bending course is provided.

Indeed, respect to the manner of the mentioned particular internal shape changes (in two different forms) in the system's fibered lightweight matrix, we have "typically non-linear strain changes in the beam height during bending" so that this non-linearly being is counted as the basic functional criterion (with its own indices) for "Resilient Composite Systems".

"If" the utilized elements in the said composite system are made of the materials, whose stress-strain diagrams within the elastic limit are partially linear (as the so-called "Linearly Elastic materials"), the system's stress-strain diagram in bending will be "non-linear (with a decreasing slope)"; however, by increasing of the endurance and strength against the mentioned internal deformities in the matrix throughout the bending course, "the decrease of the diagram's slope" will being diminished through the bending. And, in case of employing the elements made of the materials, whose stress-strain diagrams are non-linear (as the so-called "Non-linearly Elastic materials"), according to the role of each used element and its stress-strain diagram (when the element is considered by itself, out of the system), the final outcome as the stress-strain diagram of the system and its slope changes will be naturally affected. [For instance, utilizing some Polypropylene fibers instead of the fibers made of the linearly materials could lead to the comparative decrease of the said increasing slope of the system's stress-strain diagram during bending according to the case.]

Considering the texture and properties of the lightweight fibered consistent matrix (in which, the elastic strain limit ($\varepsilon_y$), the stress block indices ($\alpha$ & $\beta$) and the strain correspondent with final, complete failure ($\varepsilon_{cu}$) in compression have partially increased) and above all, respect to "*the manner of the said internal deformities* and more being of tension (stretch) in the lower parts of the beam" (which could lead to the final fracture of the beam not in the primary compressive pattern), the possibility of brittle and primary compressive fracture in the upper parts of the beam will greatly diminish, and we will have more toughness and ductility in the beam.

Indeed, in the beams made of the "Resilient Composite Systems", in bending, "the ratio of *the compressive stress leading to the supposed compressive fracture* to *the correspondent beam's strain*" and generally, "the ratio of *the maximum compressive stress in the beam in each supposed strain* to *the concurrent maximum tensile stress in the beam (in the same strain)*" (also including "the ratio *of the compressive stress leading to the supposed compressive fracture in the beam to the concurrent maximum tensile stress in the beam*") are much fewer than these ratios of the beams not having typically linear strain changes in the beam height during bending. In this way, supposed occurrence of the compressive fracture in the beam made of the "Resilient Composite Systems" in bending potentially requires considerably more strain and stress in bending compared to the similar beam not having typically linear strain changes in the beam height during bending (which naturally means the rise of the surface under the stress-strain diagram in bending up to the strain correspondent with the supposed compressive fracture in the beam).

In general, "beam fracture of primary compressive type in bending" could be occurred only when "the stress in bending required for the supposed tensile fracture occurrence in the beam" is more than "the stress in bending required for the supposed compressive fracture occurrence in the beam". The possibility of "beam fracture of primary compressive type in bending" has a direct relationship with "the ratio of *the tensile strength of the beam's tensile block* to *the compressive strength of the beam's compressive block*" multiplied by "the ratio of *the compressive stress leading to the supposed compressive fracture in the beam* to *the concurrent maximum tensile stress in the beam*".

For instance; in the beams made of the lightweight materials as lightweight concretes, the modulus of elasticity and so, "the ratio of *stress* to *strain*" and "the ratio of *the stress leading to the said supposed compressive fracture* to *the beam's strain correspondent with the supposed beam's compressive fracture*" are fewer than these ratios of the beams made of the concretes with higher densities. However, decrease of the compressive strength in the lightweight concretes leads to the increase of possibility of beam fracture of primary compressive type in bending in the beams made of the usual lightweight concretes compared to this possibility of the beams made of the concretes with more densities; whereas, in the beams made of the "Resilient Composite Systems", due to the radical decrease of "the ratio of *the maximum compressive stress in the beam in each supposed strain* to *the concurrent maximum tensile stress in the beam (in the same strain)*" (also including "the ratio *of the compressive stress leading to the supposed compressive fracture in the beam* to *the concurrent maximum tensile stress in the beam*"), this possibility is less than that of some materials with more density and compressive strength but not having non-linearly strain changes in the beam height during bending. Indeed, in the beam made of the RCS, "the ratio of *the increase of* *the maximum compressive stress* to *the increase of the maximum tensile stress* during bending" is lesser.

Generally, in the "Resilient Composite Systems", by the significant increase of the strain capability in bending, particularly within elastic extent (with non-linear strain changes in the beam height during bending), we can actually "more exploit the potential capabilities of the matrix and particularly, reinforcements in bending and tension" concomitantly. In these systems, the capability of the stresses absorption and control, the elastic strain capability and modulus of resilience in bending have been much increased.

Only employing various kinds and amounts of reinforcements as meshes or lattices, bars and polymeric or non-polymeric fibers could not lead to the mentioned favorite properties by itself. Only creating hollow pores and/or disseminating various types of elastomeric or non-elastomeric aggregates (such as Rubber, Perlite, etc) in the matrix could not result in the said particulars by itself. As well, simply reinforcing any kind of lightweight materials won't bring about the mentioned goals. To achieve the stated goals, the practical way is "creating disseminated suitable hollow pores and/or distributing appropriate lightweight aggregates in the systematically reinforced, fibered conjoined matrix".

Each component in this composition system has its important role in the ultimate result. Indeed, the components proportions and behaviors in interaction with each other bring about the above-mentioned final behavior and performance of the system. [For example, if all the said pores and/or lightweight aggregates are replaced with the Portland cement and/or sand and/or the fibered matrix used in the system (but, not including "the mentioned pores and/or lightweight aggregates"), although the compressive strength will considerably increase, but the elasticity in bending will significantly decrease, and the behavior of the system in bending will fundamentally change.]

*B. Components*

In general, the "Resilient Composite Systems (RCS)" have three necessary main elements: 1- "Mesh or Lattice"; 2- "Fibers or strands"; 3- "Matrix" with "disseminated hollow pores and/or disseminated lightweight aggregates" (in the matrix).

The last element comprises two main components;
3-a) "Disseminated hollow pores (voids)" and/or "disseminated lightweight aggregates" in the said matrix;
3-b) The cement material as a conjoined (consistent) binder. [Obviously, using the said pores and/or lightweight aggregates leads to decrease of the weight (density) according to the case.]

Naturally, the exact amount of each utilized material in these systems in each certain case depends on "numerous factors" in multilateral relationships with each other. Generally, in these integrated functioning units, the amount and manner of the mentioned components use in the organized system are always "so that" the mutual (reciprocal) interactions among the components finally lead to the "typically non-linear strain changes in the beam height during bending" (as the "basic functional character" of these systems, with its specific testable criteria and indices) and fulfillment of the functional specifications of the system in practice. [As well, the said main functional character is *much so that* we cannot use the relations and equations based upon the basic assumption of "linearly being of the strain changes in the beam height during bending" to realistically analyze the behavior of these systems.]

Here, we want to discuss partially more about the components of the RCS:

*1) Mesh or Lattice:* The used meshes or lattices could be made of the materials such as steel, polymeric or composite materials, etc. Anyway, as a rule, the modulus of elasticity and elastic strain limit ($\varepsilon_y$) "in tension" of the mesh or lattice employed in the said composite system is necessarily more than those of "the fibered matrix used in the composite system also having lightweight aggregates and/or hollow pores (but not together with lattices or meshes as tensile reinforcements)". (Naturally, the kind, dimensions, shapes and directions of the utilized meshes or lattices could be different according the case.) [In theory, if the used fibered matrix and employed mesh in the system concurrently reach to the elastic strain limit ($\varepsilon_y$) in bending (together with each other), we will get access to the most use of the potential capacities of the materials in bending. In this case, the fracture toughness will decrease. It is clear that; according to various parameters as the application case, pattern of fracture, etc, any "probable" employment of the additional and accompanying elements (such as the supplementary reinforcements in the more in-tension areas to increase the resistance and fracture toughness, etc) together with the mentioned system could be taken into consideration. (However, these various elements are not counted as the necessary components of the systems generally called as "Resilient Composite Systems".)]

*2) Fibers or Strands:* The used fibers or strands could be kinds of flexible polymeric or non-polymeric fibers (such as Polypropylene fibers, Polyester fibers, and steel fibers). As a rule, the modulus of elasticity and elastic strain limit ($\varepsilon_y$) "in tension" of the fibers employed in the said composite material are necessarily more than those of "the matrix used in the composite material also having lightweight aggregates and/or hollow pores, but without fibers". [As well, the length of the fibers should be at least more than the longest length of the existent pores or aggregates, when they are in their maximum stretch in the system (in the strain correspondent with the beam's final failure).]

*3) "Matrix" with the Suitable Hollow "Pores (Voids)" and/or "Lightweight Aggregates" in its Context:* About the matrix of the system, as its binder with the expedient particulars as consistency, flexibility, etc, it should be also mentioned that:

3-1- When we use the term of "Cement Material", it means the conjoined (consistent) binder employed as the context of the system generally called as Composite (in its broad meaning). In the "Resilient Composite Systems", we could use a wide range of cement materials such as: net (pure) Portland cement plus water, the composition of the Portland cements with Pozzolanic materials plus water, the composition of Pozzolanic materials and lime plus water, polymeric cements, etc.

Naturally, no gravel is employed in the matrix of the "Resilient Composite Systems". As well, here, sand is not a necessary element. [If sand is probably used in the system, it should be "fine" and "well conjoined to the cement material". Otherwise, it will dramatically result in serious disturbances in the behavior and performance of the matrix and system and bring about the problems such as; falling of the modulus of resilience and bearing capacity in bending, increase of the possibility of brittleness and non-security being of the fracture pattern, etc. Generally, it is better that no sand or any other non-cement (non-active) material is used in the matrix "if possible". Nonetheless, "if" because of any reason, the non-cement materials with high fineness are utilized in these systems and the cement material of the system is "the mixture of Portland cement and water" or "the mixture of Portland cement and Pozzolanic materials and water" or "any other cement material including the C-S-H crystals", it is possible by reducing the ratio of the cement materials to the water (for instance, to less than 0.4), the consistency of the matrix is comparatively improved. Meanwhile, appropriately employing some expedient Pozzolanic materials such as micro Silica fume could lead to creation of some C-S-H crystals with smaller sizes in the matrix (also among the bigger crystals of C-S-H and within the interfaces existing between the cement and non-cement materials) and brings about partially more consistency and behavior in the matrix.]

3-2-1- The mentioned hollow pores disseminated in the matrix could be created by various methods, such as some common methods used in making the gas bulbs in the cellular or foam concretes, etc.

3-2-2- Lightweight aggregates disseminated in the matrix could be kinds of polymeric and/or non-polymeric aggregates (such as the beads or particles of Rubber, Plastic, Polypropylene, Expanded Polystyrene, Perlite, Vermiculite, etc). As a rule, the density and compressive modulus of elasticity of the lightweight aggregates employed in the said composite system are necessarily fewer than those of the "the fibered matrix used in the composite system, but without lightweight aggregates and/or hollow pores". The employed lightweight aggregates' stress-strain diagrams in compression, at least in all strains up to "the strain correspondent with the compressive strength" of "the used fibered matrix but without lightweight aggregates and hollow pores", are necessarily under the stress-strain diagram in compression of "the fibered matrix without lightweight aggregates and hollow pores". [As it was pointed before; here, the general term of "lightweight aggregate" has a broad meaning, also including the polymeric and non-polymeric beads or particles.]

It should be also mentioned that; the main role of the appropriate lightweight aggregates using in the structure (make) of this system is to create the disseminated expediently flexible regions in the matrix.

Generally, flexibility has a wide and partial meaning. Even the materials with comparatively low flexibility compared to some other materials as the typical elastomeric materials such as rubber could be also used as the lightweight aggregates in this system considering the mentioned requirements. (Even, after employing them in making the system, some of these materials may be crashed in the system under high strains in bending. However, the main role of them in forming of the system and getting access to the mentioned structure has been already fulfilled.)

Naturally, employing as much as finer pores and/or lightweight aggregates and employing the lightweight aggregates with more elastic strain limit ($\varepsilon_y$) and "modulus of elasticity" in compression (but still lower than that of the fibered matrix used in the system) could finally lead to better behavior and more endurance limit and modulus of resilience in compression and bending in the system; nevertheless, using elastomeric particles or beads as lightweight aggregates is not "necessary" (inevitable) for getting access to the so-called "Resilient Composite Systems" with the mentioned specifics. (Although the properties of the used aggregates as their modulus of elasticity, permeability, durability, etc are all effective in the final outcome, but any elasticity of them is not the main cause of the system's high modulus of resilience in bending.)

It is clear that; presence of enough expediently flexible regions in the used matrix is another necessary condition to attain the "Resilient Composite Systems". In this regard, the percentage of the total space occupied by the employed lightweight aggregates and/or pores in the used matrix with the lightweight aggregates and/or hollow pores (but not together with the fibers) could be an index in its turn.

**-** As it was stated; without any elasticity in the "matrix" (also before employing the fibers or strands) getting access to the "Resilient Composite Systems" would be impossible; however, elasticity, similar to the flexibility, is a partial property. For instance, the ratio of "the elastic strain limit ($\varepsilon_y$) in compression" to "the strain correspondent with the compressive strength" in the used matrix without lightweight aggregates and/or hollow pores (and not together with the fibers or stands) could be counted as an index in this regard. [Naturally, the certain percents related to these ratios and percentages could be, according to the case, established considering more detailed studies in the field.]

In general and similar to the other components (as meshes or lattices and fibers or strands), amount and properties of the employed matrix (with hollow pores and/or lightweight aggregates) in each case should be so that the stated requirements for the other components in the system and "the mentioned functional criteria for the system" are finally fulfilled.

Furthermore, contrary to some other composites and so-called elastic or resilient concretes and the like, here, employing some expensive polymeric cement materials is not a "necessary" and inevitable condition or specification to achieve the stated specifics for the system. (For instance, only the common mixture of Portland cement and water or preferably, the mixture of Portland cement, some appropriate Pozzolanic materials and water could be also used as the cement material of the system if needed.)

*C. More Explanations about the RCS*

"The strain changes in the beam height during bending" is counted linearly in "the Basic Kinematic Assumption of the Flexure Theory" in the Solid Mechanics. This fundamental and primary assumption and its derived relations are the base of many employed equations in the field.

For instance, many usually employed equations to calculate the quantities such as modulus of resilience, ultimate strength moment in beams, and equilibrium reinforcement amount ($\rho_b$) (in order to attain the beams with the fracture pattern of secondary compressive as a secure fracture pattern in bending) are based on this basic assumption.

Considering the mentioned basic assumption, there are some fundamental limitations in raising the modulus of resilience and ultimate strength moment "in bending" and employing tensile reinforcements in beams.

Particularly, in the lightweight beams having low compressive strength and modulus of elasticity, these limitations are more sensible.

Respect to the basic kinematic assumption of the flexure theory, less compressive strength leads to more possibility of "the beam fracture of primary compressive type". And, strengthening the so-called tensile block in the beam to increase modulus of resilience and bearing capacity in bending (for instance, by employing more tensile reinforcements), without expediently strengthening the compressive block concurrently, increases the possibility of the beam fracture of primary compressive type as a non-secure pattern of beam fracture in bending. [For instance, in the low height reinforced beams (as slabs) with comparatively low weight and compressive strength, the equations for calculation of the required compressive reinforcements used to concurrently strengthen the compressive block could result in very huge amounts as the required compressive reinforcements to get access to the beams with high bearing capacity and secure fracture pattern.]

It is clear that; only reinforcing the materials with bars, meshes and fibers to increase the modulus of resilience in tensile and bending is an experienced method, and it is not novel. [*For instance,* "Ferrocements", Fibered Concretes, and Lightweight Concretes have been also discussed in "ACI 544", ACI 549, and ACI 523 respectively. (As well, the cellular concretes, the lightweight concretes containing Expanded Polystyrene beads and other kinds of the so-called insulating (insulant) lightweight concretes have been also pointed in ACI 523.1R-92.)] Nonetheless, only employment of various types and amounts of the tensile bars, meshes and fibers to increase the modulus of resilience in bending could not concomitantly lead to having a material with "less possibility of the beam fracture of primary compressive type" and "significantly less weight" accompanied by typically non-linear strain changes in the beam height during bending altogether.

Conversely, considering the basic kinematic assumption of the flexure theory, only increasing the tensile strength and more employment of the tensile reinforcements in any type (such as fibers, various meshes or lattices, bars, etc) to raise the modulus of resilience, by itself, could lead to "increase of the possibility of beam fracture of primary compressive type". And, raising the height of the beam or raising the compressive strength by increasing the density to decrease the possibility of primary compressive fracture in the beam could naturally result in higher weight in the structure.

According to a general rule in the "Solid Mechanics" ("Strength of Materials"), *fewer density results in less compressive strength*. There is a known relationship between density and compressive strength in the solid materials, also shown by a particular diagram (with the lessening slope).

Thereby, decreasing the density of a material (for instance, by creating disseminated pores and/or by disseminating the lightweight aggregates in that material) leads to decrease of the compressive strength. As well, compressive strength has a direct relationship with tensile strength and then, modulus of resilience in bending. Meanwhile, also considering the basic kinematic assumption of the flexure theory, decrease of the density could bring about more possibility of the beam fracture of brittle and primary compressive type in bending. [The recent effects are due to more possibility of the matrix rupture and fracture under the applied stresses.]

It is worthy of saying that; with the basic assumption of "linearly being of the strain changes in the beam height during bending", any effect of creating the hollow pores and or employing lightweight aggregates in a solid material (such as the concrete and the like to decrease the density) in rise of "the ratio of *strain* to *stress* (which means lower modulus of elasticity) and decrease of "the ratio of *the compressive stress leading to the supposed fracture in the compressive block of the beam* to *its correspondent strain in bending*" cannot, by itself, neutralize "the crucial effect of the compressive strength decrease" due to the density reduction. Thereby, occurrence of fracture in the compressive block of the beam requires "less energy through bending" and thereby, "the possibility of beam fracture of primary compressive type in bending" increases.

Anyway, despite the mentioned subjects, here, we have tried to present a simple and practical method to: 1- significant increase of the modulus of resilience in bending; 2- considerable decrease of the possibility of the beam fracture of brittle and primary compressive type in bending; 3- large reduction of the weight. Indeed, through making a particular integrated functioning system, for the first time, the afore-said (paradoxical) items have been concomitantly fulfilled "altogether". [As well, there is no record of employing any kind of "the lightweight fibered ferrocements" and the like "as the material with typically non-linearly strain changes in the beam height during bending, having its own specifications and functional criteria". Naturally, here, the words as "significant" and "considerable" have their own indices, which are stated in the testable criteria of the compound materials as the "Resilient Compound Systems".] Meanwhile, through employing the presented method, "the strain changes in the beam height during bending" will be typically non-linear (which has its own testable criteria and indices).

As it has been pointed before; the mentioned internal shape changes (deformities) in the ameliorated matrix during the bending in the "Resilient Composite Systems" are so that "the strain changes in the beam height during bending" is non-linear, and this non-linearly being could be counted as the basic functional criterion for the 'Resilient Composite Systems' in its turn.

It is worthy of saying that; even in the usual reinforced concrete beams, the strain changes in the beam height during bending is not "exactly" linear. As it has been also mentioned in the texts of "Strength of Materials" [3], even in the beams made of the usual reinforced concrete, "in fact", the strain changes in the beam height during bending is fairly non-linear. Nonetheless, this non-linearly being is *little so that* it is ignored in practice. In the usual reinforced concrete beams, the common used relations & equations are based upon the assumption of "linearly" being of the strain changes in the beam height during bending. The said main assumption, considering its derived trigonometric equalities and proportions, is the base of some common used equations in practice. For instance, the usual calculation of the modulus of resilience (within the elastic limit), the bearing capacity and equilibrium steel ($\rho_b$) in the usual reinforced concrete beams are all based upon this primary assumption of bending in the solid materials. As well, despite the exact actual pattern of the elastic deformities in the usual reinforced concrete beams during bending, they are not counted as "the materials having typically non-linear strain changes in the beam height during bending".

When we clearly count a material as the material having "typical non-linear strain changes in the beam height during bending", it specifically means that; "the strain changes in the beam height during bending" is *non-linear so that* the mentioned basic assumption of bending and the equations based upon the mentioned assumption (which are "for instance", commonly used to calculate the ultimate strength moment and modulus of resilience in bending in the usual reinforced concrete beams) do not, even roughly, hold true.

Thus, we need different equations (with different basic assumption of bending) to analyze the behavior of the system and calculation of the modulus of resilience, bearing capacity, etc in the structure. Indeed, contrary to the solid materials (such as the usual reinforced lightweight concretes and the like) in which, "the strain changes in the beam height during bending" is not "typically" non-linear, in the said composite material, the error resulted from the existent difference between the "actually" non-linearly being of the strain changes in the beam height during bending and the said "assumption" is *large so that* we cannot utilize the equations based upon the said assumption to even partially "realistically" analyze the behavior of the mentioned material in bending.

This could be, in its turn, a "basic functional criterion and index" to differentiate the materials counted as "the materials having the so-called linear strain changes in the beam height during bending" from the materials counted as "the materials having the so-called typically non-linear strain changes in the beam height during bending". Generally, when we specifically count a material as a material having "typically non-linear strain changes in the beam height during bending", we should also be able to present in practice that; "the basic kinematic assumption of the flexure theory and its derived relations & equations" do not even roughly hold true in that material (which could be determined by the related indices and practical criteria).

Here it is worth mentioning that; contrary to some previously mentioned issues in the field, the meaning of the "non-linearly being of the strain changes in the beam height during bending" (as the basic assumption of the "Flexure Theory" in the "Solid Mechanics") is naturally different from the "non-linearly being of the stress-strain diagram during bending" (as the behavior description of some materials generally called as "Non-linearly Elastic Materials" in Mechanics). As it was discussed before; employing the non-linearly materials in making the "Resilient Composite Systems", as their components, is not a necessary (inevitable) condition to achieve the basic functional character of these systems as the "non-linearly being of the strain changes in the beam height during bending". As well, by meticulously employing some particular non-linearly materials having "increasing slope of the stress-strain diagram in bending" as the components of the "Resilient Composite System", it is possible to get access to the type of the "Resilient Composite Systems" with partially linearly stress-strain diagram in bending; while, the strain changes in the beam height during bending is still non-linear. [Even, sometimes, employing some particular methods to drive (impel) the pattern of "the stress-strain diagram during bending" to more "non-linearly being" may, according to the case, lead to decrease of the modulus of resilience in bending. As an example, in a structure as the so-called "Linearly Elastic Structure" (in bending), replacing "some of the used Linearly Elastic Materials (as the system's components) having more modulus of resilience in bending" with "some of Non-linearly Elastic Materials having less modulus of resilience" could drive the structure's stress-strain diagram to non-linearly being (by decreasing the stress-strain diagram's slope) and reduce the modulus of resilience in bending concurrently.]

Moreover, there are various ways to significantly increase the modulus of resilience in bending; but, not all of them drive the pattern of "the strain changes in the beam height during bending" to the so-called "typically non-linearly being" (with its own criteria and indices).

Hence, in the "Resilient Composite Systems", more strain capability particularly within the elastic limit (with the said particular pattern) brings about better distribution of the stresses and strains and more benefiting the in-tension reinforcements' potential capacities. This also means delaying in establishment and outspread of the cracks and effective damages in the matrix. As well, the energy absorption and energy reserving capacities in this system will be high.

Thus, respect to providing the appropriate strength reserving and high modulus of resilience, together with less possibility of the beam fracture of primary compressive type in bending in less weight (density), we could constructively access to the "high capacity of load bearing in bending" despite the low weight and dimensions.

*D. Why are These Systems Called as "Composite"?*

The term of "Composite" in general has a broad meaning. Generally, when we have "more than one component" and the existent "interactions" among the components in the *compound* lead to a "unique functioning system", we can call that system as the so-called "Composite" in its "broad meaning". For example, even usual concrete, which is only composed of the Portland cement, water and aggregates, could be counted as a "Composite" in its wide meaning. Thereby, the materials as Ceramic and Polymeric Materials in general are not the "necessary" (inevitable) specific components for any system called as "Composite". [In this way, any kind of concrete is also a kind of "Composite" "in its broad meaning"; but obviously, any kind of "Composite" is not a type of Concrete.]

However, the term of "Composite" in its "specific meaning" indicates the systems, as the integrated functioning units, generally consist of: 1- Matrix; 2- Fibers or Strands; 3- Mesh or Lattice, as the "three necessary main components" in each material or system called "specifically" as "Composite" system (in its specific meaning).

In many times, the mentioned matrix is made of the expedient Ceramic or polymeric materials in general; but these materials are not always the necessary, inevitable materials for counting a compound as the "Composite" in its general and also specific meanings.

*E. The General Structural Particulars and Functional Criteria as the Necessary Specifications of the Compound Materials Generally Called as "Resilient Composite Systems":*

Considering the mentioned items, we can present some necessary general structural particulars and functional criteria and indices for the systems generally called as "Resilient Composite Systems".

*1) General Structural Criteria:* Presence of the necessary and basic components altogether as; *mesh or lattice* (in any kind, but necessarily having the stated general requirements); flexible *fibers or strands* (in any kind, but necessarily having the stated general requirements); cement material (as the conjoined binder in any kind, but necessarily having the stated general requirements); disseminated hollow pores *and/or* lightweight aggregates (in any kind, but necessarily having the stated general requirements) in the matrix. [Here, the general term of "lightweight aggregate" has a broad meaning, also including the polymeric and non-polymeric beads or particles.]

As it has been mentioned before; naturally, the exact amount of each utilized material in these systems in each certain case depends on "numerous factors" in multilateral relations with each other. Generally, in these integrated functioning units, the amount and manner of the mentioned components use in the organized system are always *so that* the mutual (reciprocal) interactions among the components finally lead to the *typically non-linear strain changes in the beam height during bending*" (as the "basic functional character" of these systems, with its specific testable criteria and indices) and fulfillment of the practical functional specifications of the system. [As well, the said main functional character is *much so that* we cannot use the relations & equations based upon the basic assumption of "linearly being of the strain changes in the beam height during bending" to realistically analyze the behavior of these systems.]

*2) Functional Criteria (Required Specifications):* As it was stated; having "typically non-linear strain changes in the beam height during bending" is the "basic functional criterion" of the system. This in practice means; "the strain changes in the beam height during bending" is *non-linear so that* "the equations based upon the basic assumption of linearly being of the strain changes in the beam height during bending" (which are for instance, commonly employed to calculate the ultimate strength moment and modulus of resilience in bending in the usual reinforced concrete beams) do not even roughly hold true. Naturally, the mentioned "*typically* non-linear being" has its own practical criteria and indices. Thus, the composite systems made by this method have also their own criteria.

Generally, the practical and testable functional criteria are founded on the difference between "the actual behavior of the beam made of the said composite systems" and "the calculated nominal quantities derived from the relations & equations based upon the basic kinematic assumption of the flexure theory" (in which, the strain changes in the beam height during bending are assumed linear). When the mentioned difference is more than a certain percentage, we count the material as the material having the so-called typically non-linear strain changes in the beam height during bending. As well, according to the case, the manner and amount of this difference between "the actual behavior of the said materials" and "their expected behavior in bending upon the basic kinematic assumption of the flexure theory" could be so that the testable functional specifications as the specific functional criteria are practically fulfilled.

For instance and as the "Practical Functional Criteria":
*a)* The "actual" ultimate strength moment ($M_e$), "cracking moment ($M_{cr}$) and especially, *elastic strain limit* ($\varepsilon_y$)" "in bending" in the beams made of the mentioned composite systems are significantly more than the "nominal" quantities of the ultimate strength moment ($M_n$), cracking moment and elastic strain limit in bending "calculated by the equations based upon the basic kinematic assumption of the flexure theory"; [As it was mentioned before; naturally, the certain percents related to these ratios and percentages could be, according to the case, established considering more detailed studies in the field.]
*b)* Increase of the used tensile reinforcements as the bars, lattices, etc in the beams made of the mentioned systems significantly more than the so-called equilibrium reinforcement amount "calculated by the equations based upon the basic kinematic assumption of the flexure theory" cannot bring about the primary compressive fracture pattern under the in-bending loadings (more than the actual ultimate strength moment of the beam).
- The above-mentioned functional criteria are "concomitantly" fulfilled in the said composite systems (altogether).

Thereby, we could name the mentioned made materials as the "Resilient Compound Systems" or preferably, "Resilient Composite Systems (RCS)".

### III. "ELASTIC COMPSITE REINFORCED LGHTWEIGHT CONCRETE (ECRLC)" AS A TYPE OF THE RESILIENT COMPOSITE SYSTEMS (RCS)

Here, the "Resilient Composite Systems" (with the mentioned properties and criteria), whose cement materials include the "C-S-H (Calcium Silicate Hydrate) crystals" have been named as the "Elastic Composite, Reinforced Lightweight Concrete (ECRLC)".
For instance, the composition of the Portland cement and water, the Portland cement and water and Pozzolanic materials, and lime and Pozzolanic materials all are the cement materials comprise the C-S-H crystals.

Respect to the required general structural properties and specific functional criteria mentioned for the "Resilient Composite Materials" and considering the application case, the content of the consistent cement material is as much as needed to provide the necessary bonding and behavior in the system. Many times, we need more cement in this structure compared to some usual reinforced concretes. [Hence, employing the Portland cement type II could be, "according to the case", taken into consideration.]

Besides, there are suitable pores and/or appropriate lightweight aggregates (such as the Expanded Polystyrene beads [1], etc) and the fibers or strands (such as the Polypropylene fibers, Polyester fibers, etc) in the consistent matrix. The employed mesh or lattice could be also kinds of the welded steel wire meshes (lattices) in expedient shapes and dimensions or other types of meshes or lattices. Anyway, also the general structural properties and requirements mentioned for the particular compound materials named as the "Resilient Composite Systems" should be considered in the kind and composition of the components comprising the ECRLC as the final functioning system. As it was mentioned before; using the said hollow pores and/or lightweight aggregates leads to decrease of the density. In this way, we could also get access to the so-called (thermal) insulating materials according to the case.

Existing of the said organized reticular structure in the context of the system could assist to better control of the possible cement materials contractile stresses and comparatively useful accumulation of the mentioned stresses in the piece; just as, in appropriate conditions, this matter could, in its turn, lead to partial increase of the system's strength in tension and bending and could also affect its ductility.

Generally, fibered being of the said fibered lightweight matrix, adhesive being of the conjoined cement material (considering the used components type and amount) and extensive surface of the utilized reinforcements (in form of the mesh or lattice, for example with the connected, perpendicular longitudinal and transverse components) are all impressive in appropriate involvement of the reinforcements in the mentioned fibered lightweight matrix. Meanwhile, regarding the general properties of the said fibered lightweight matrix, "the system's reticular structure" and particularly, the utilized fibers with suitable involvement in the bonding matrix can lead to useful control of the shrinkage stresses and the like. (And, good control of the shrinkage effect could, in its turn, assist to increase of involvement of the used fibers and lattice or mesh in the consistent matrix.)

In the ECRLC and in general in the RCS, matching behavior of the in proportion, complementary components in appropriate interaction with each other could be counted as the factor of achieving the mentioned specifics in the system. In this way, the elasticity and modulus of resilience and bearing capacity in bending in ECRLC are not only significantly more than those of the usual reinforced lightweight concretes with the same densities and having similar reinforcements, dimensions and cement materials, but also more than those of some usual reinforced concretes with significantly higher densities and compressive strengths (of the concrete) and having similar reinforcements and dimensions. Besides, respect to the mentioned issues about the system's fracture pattern in bending, here, the usual calculation of the equilibrium steel amount for attaining the low-steel bending sections with secure fracture pattern (of secondary compressive type) and the related usual limitations do not become propounded.

*An Instance of the Lightweight Concrete that Could be Used in Making the ECRLC*

Here, only as an instance, we have pointed to a special type of lightweight concrete that could be methodically reinforced in the framework of the particular composite system called as ECRLC.

Among the said utilized fibered lightweight concrete's properties we could point to: the appropriate ratios of "the elasticity modulus, tensile, and shearing strengths" to "the compressive strength"; the appropriate ratios of "the surface under the stress-strain diagram and also the strength in the elastic limit" to "the ultimate strength" (particularly, with respect to the lightweight concrete's density); the appropriate ductility; highly being of the strain correspondent with the compressive strength; significantly highly being of the strain correspondent with the final, complete failure ($\varepsilon_{cu}$) in compression; the appropriate fracture toughness (with high α and β stress block indices) and so, the non-brittle fracture pattern as a kind of being compressed in high in-compressive loading (partially similar to the so-called paste-form materials) instead of the typical extensive shattering (as the usual pattern of crashing in the usual concretes). [It is worthy of mentioning that; regarding the significant toughness and occurrence of a type of "gradually collapsing (being compressed)" in this kind of lightweight concrete (especially of the fibered type), here, the subject of "the strain correspondent with the final, complete failure ($\varepsilon_{cu}$) in compression" in its current meaning in the usual concretes ("as a certain, exact quantity") could lose its particular point.] (Fig. 1)

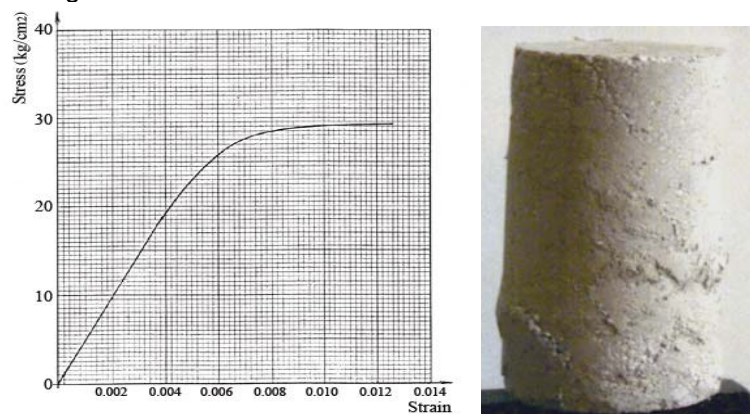

Fig. 1. The stress-strain diagram of in-compressing loading of the pointed lightweight concrete containing the Expanded Polystyrene (EPS) beads as the lightweight aggregates.
- Oven-dry density = 600kg/m$^3$; f´$_c$ = 29.5kg/cm$^2$ in 28 days; micro Silica fume / Portland cement (Type II) = 8.5%; W / C+S = 0.425 (with using Lignosulfonate as a common "plasticizer" and retardant); monofilament Polypropylene fiber (Denier: 3) = 1.1% of volume of concrete. (In this study, gravel and sand have not been employed in the used fibered lightweight concrete.)
 - Here, the strain correspondent with the final, complete failure ($\varepsilon_{cu}$) in compression in its current meaning in the usual concretes and as a certain exact quantity cannot be determined. [$\varepsilon_{cu}$ ↑]

However, it should be mentioned that; although, reaching to the said properties in a lightweight concrete could be important in its turn, but without systematically employing and expediently reinforcing the afore-said special lightweight concrete and the like in a methodical framework in order to achieve the particular structure called as ECRLC (as a type of the "Resilient Composite Systems" with the stated structural properties and specific functional criteria), getting access to the peculiarities of the systems generally called as "Resilient Composite Systems" (such as the significant specific modulus of resilience and resistivity in bending) would not be conceivable. None of the considerable above-mentioned properties, by itself, means highly being of the modulus of resilience in bending necessarily (also "compared to" some reinforced lightweight concretes with considerably more densities and compressive strengths of the concrete and having the similar reinforcements and dimensions).

By usually reinforcing the above-said fibered lightweight concrete with the tensile bars or even meshes, but *not "methodically"* in the framework of the particular reinforced system named as ECRLC, we cannot finally attain the specifics and particular advantages of the ECRLC.

Only by using all of the stated components in the framework of the special system called as ECRLC (as a type of the RCS), we can achieve the said significantly high modulus of resilience in bending and other particulars and preferences of the ECRLC; not only in comparison with that of the usual reinforced lightweight concretes with the same densities and having similar reinforcements and dimensions, but also in comparison with that of some usual reinforced concretes with considerable higher densities and compressive strengths (of their concretes).

## IV. REVIEW OF SOME EXPERIMENTS, AND MORE DESCRIPTION ABOUT ECRLC

Regarding the mentioned subjects and claims, the results of some actual performed loadings of the slabs made of the above-said "Elastic Composite, Reinforced Lightweight Concrete" ("as a type of the said composite structures with the stated general structural properties and specific functional criteria") in a manner similar to the ASTM E 72 are considerable.

In the analysis of the mentioned "in-bending structures" (with equations related to the usual reinforced concretes to calculate the nominal ultimate strength moment ($M_n$) by the method called as "Ultimate Strength" [4], [5] (which is also based upon the *kinematic assumption of flexure theory*), the "nominal" ultimate strength moment ($M_n$) has been much less than the "actual" amount of the beam's ultimate strength in bending in practice ($M_e$).

*Even when the concrete's "compressive strength" in the related equations ("based upon the said basic kinematic assumption of flexure theory"), from mathematical point of view, is inclined toward extreme (∞) and the stress block height has been supposed equal to zero, the $M_e$ has been obviously higher than $M_n$.*

Particularly, in calculating the nominal amounts of the "cracking moment ($M_{cr}$)", "elastic strain limit ($\varepsilon_y$)", and "modulus of resilience" ($u = \frac{1}{2} \sigma_y \cdot \varepsilon_y$)" [3] in bending, the "actual" amounts (especially, the actual amount of the "strain elastic limit ($\varepsilon_y$) in bending") have been significantly higher than the nominal amounts derived from the common relations & equations employed for analyzing the usual reinforced concrete beams (based upon the basic assumption of "linear being of the strain changes in the beam height during bending").

In fact, in respect of the "actual" behavior of the system throughout the bending, despite much increase of the applied tensile forces in the slab in bending course from what is called as "compressive block strength" (calculated by the usual relations & equations based upon the said basic assumption), the beam's strain in bending is still being continued up to reaching to the higher amounts of strain.

Meanwhile, after passing through the elastic and plastic stages, occurrence of the final bending fracture in the slab has been resulted from the gradual occurrence and deepening of the cracks in the under grater stretch (tension) layers in the beam with a non-brittle pattern. As well, in spite of significantly higher being of the utilized tensile steel amount in the mentioned slab than the equilibrium steel («$\rho_b$», calculated according to the common relations & equations related to the usual reinforced concrete beams ($\varepsilon_{cu} = 0.003$), also based upon the basic kinematic assumption of flexure theory) [4], [5], the fracture patterns in the slabs have not become primary compressive type and brittle. (Figures 2.1. 2.2, 2.3, 2.4, 2.5, 2.6, and 2.7.])

[Naturally, if there was no "unwanted, asymmetrical primary deflections" in the slabs in the tested slabs before the loadings (shown in the Fig. 2-3, and due to some unwanted difficulties in the used inappropriate wood molds in making the slabs) and if the loading was performed "exactly" according to the ASTM E 72 standard, the cracks would be more "disseminated" and the elastic strain capability, modulus of resilience and loading capacity would be "*considerably higher than the obtain results*" in these tests. Meanwhile, the existent unwanted defects in the welding points of the employed welded steel wires lattice (mesh) could naturally bring about decrease of the elastic strain limit ($\varepsilon_y$), the modulus of resilience and bearing capacity in bending in its turn.]

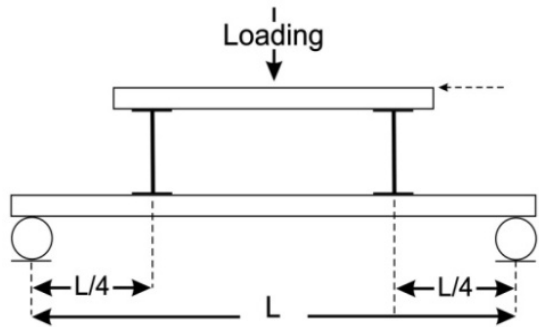

Fig. 2.1. The manner of "in-bending" loadings of the instances of the mentioned slabs (made according to the system as the ECRLC) in a manner similar to the ASTM E 72 standard

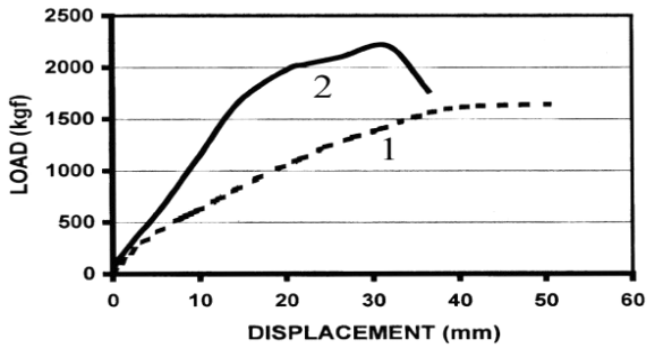

Fig. 2.2. The load-displacement diagram of the "in-bending" loadings of two instances of the mentioned slabs (made according to the system as the ECRLC) "with and without the supplementary bars" in a manner similar to the ASTM E 72 standard
- The first slab (scattered line) has no supplementary (supportive) bar, and the second (filled line) has the supplementary (supportive) bars.
[The stated results have been obtained despite the presence of "the unwanted primary, asymmetrical deflections" in the tested slabs. The "structure", "related dimensions and quantities", and "amounts of the mentioned deflections in the (concave) slabs before loading" have been shown in the figure 2.3. (Naturally, previous existent deflections in the slabs increase through the loading.)]

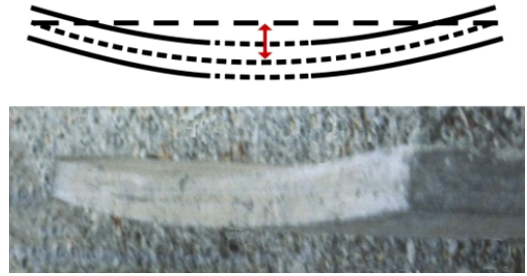

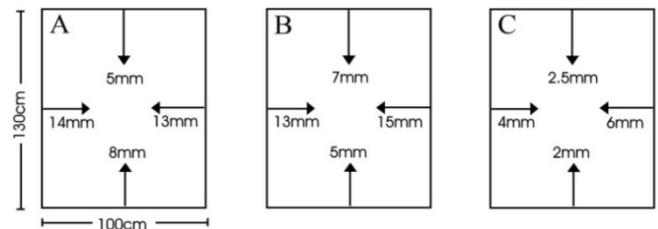

Fig. 2.3. The amounts of the unwanted, asymmetrical primary deflections in the centers of the tested slabs' sides (concerning horizon line) before loading (mm)
A) With the supplementary (supportive) bars (for the in-compressing loading); B) Having the supplementary (supportive) bars (for in-bending loading); C) Not having the supplementary (supportive) bars (for in-bending loading)

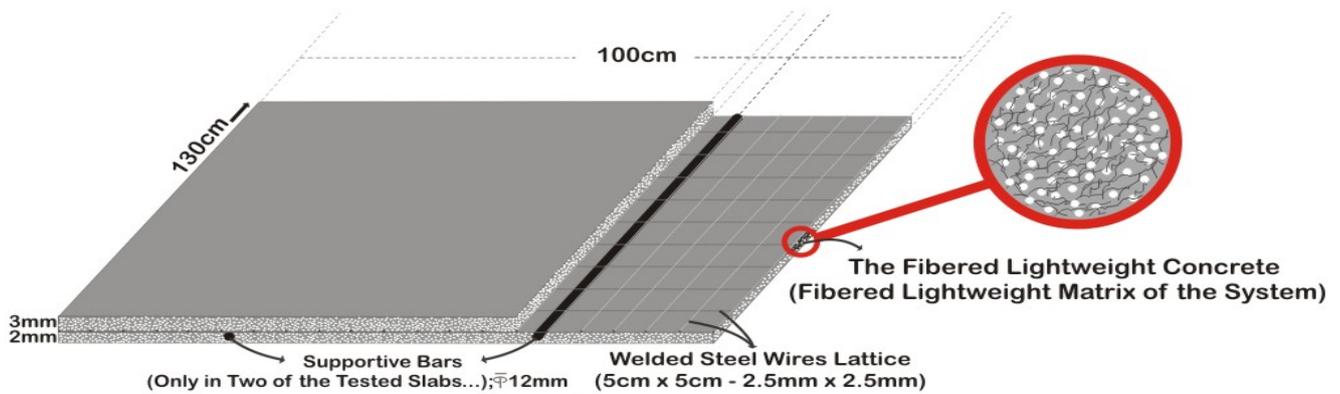

Fig. 2.4. The structure and related dimensions and quantities in the tested slab (made according to the mentioned system as the ECELC) with the supplementary bars [also including the simple recipe (mixture plan) of the used fibered lightweight concrete (as the fibered lightweight matrix) and the said structure in detail for any repetition of the tests]
● Dimensions: L = 120cm, h = 5cm, b = 100cm
● Welded steel wires lattice (mesh) (made of cold-drawn steel wires, and having some unwanted defects in the welding points):
5cm×5cm – 2.5mm×2.5mm
$f_{y1 (Mesh)}$ = 4672kg/cm$^2$, $A_{s1 (Mesh)}$ = 0.98cm$^2$, $d_{1 (Mesh)}$ = 3cm, $E_s$ = 2×10$^6$kg/cm$^2$
[The lattice's longitudinal steel wires were "on" the lattice's transverse steel wires (100cm) at the time of the in-bending loadings.]
● Supplementary steel bars (as the additional, accompanying element "in two of the tested slabs")
$f_{y2 (Bar)}$ = 4400kg/cm$^2$, $A_{s2 (Bar)}$ = 2.26cm$^2$; $d_{2 (Bar)}$ = 3.9cm, $E_s$ = 2×10$^6$kg/cm$^2$
● Fibered lightweight concrete (as the "fibered lightweight matrix" having the Expanded Polystyrene beads (as the lightweight aggregates):
$f'_c$ = 64kg/cm$^2$, $f_r$ = 34.5kg/cm$^2$, $f_{ct (Brazilian Method)}$ = 14.5kg/cm$^2$, $E_c$ = 4×10$^4$kg/cm$^2$
● Recipe ("Mixture Plan") of making the used special fibered lightweight concrete, which methodically reinforced in the particular framework of the shown system in this study: *Portland Cement (Type II) + Micro Silica Fume (8.5% of Cement Materials) = 675kg/m$^3$; W / C+S = 0.425 (with* using Lignosulfonate as a common "Plasticizer" and Retardant); monofilament Polypropylene Fibers (denier: 3) = 12.6kg/m$^3$ ("with two different lengths": 2 portion in 12mm and 1 portion in 6mm, which could be blown and mixed with "scanty" amount of the micro Silica fume for better separation and distribution of the fibers); Expanded Polystyrene (EPS) beads ($D_{50}$ = 3.2mm) up to 1m$^3$.

- In this study, gravel and sand have not been employed in the used fibered lightweight concrete. [Generally, if sand is probably employed in these systems, it should be "fine" and especially, "well conjoined to the cement material". Otherwise, it will dramatically result in serious disturbances in the behavior of the system and bring about the problems such as; considerably falling of the modulus of resilience and bearing capacity in bending, brittleness of fracture pattern, etc. (In general, it's better no non-cement material (as the sand) be used in the matrix if possible.)]
- Curing of the fibered lightweight concrete (fibered lightweight matrix) used in the slabs has been performed via the so-called Membranous Method (for 30 days).
- The tests have been done about 90 days after making the slabs, and the compressive strength of the used fibered lightweight concrete (also cured by the membranous method) has been simultaneously measured with loading of the slabs 90 days after making the samples.
- Oven-dry density & drying shrinkage (90 days) of the fibered lightweight concrete reinforced in the framework of the system in this instance have been respectively about 835kg/m$^3$ and less than 0.015.
- [It is worthy of mentioning that; generally, the presence of supplementary bars as shown here is not a necessary condition to count a system as the RCS or ECRLC (as a type of RCS). However, here, considering the performed studies, these additional elements are used accompanied by the system named as ECRLC.]

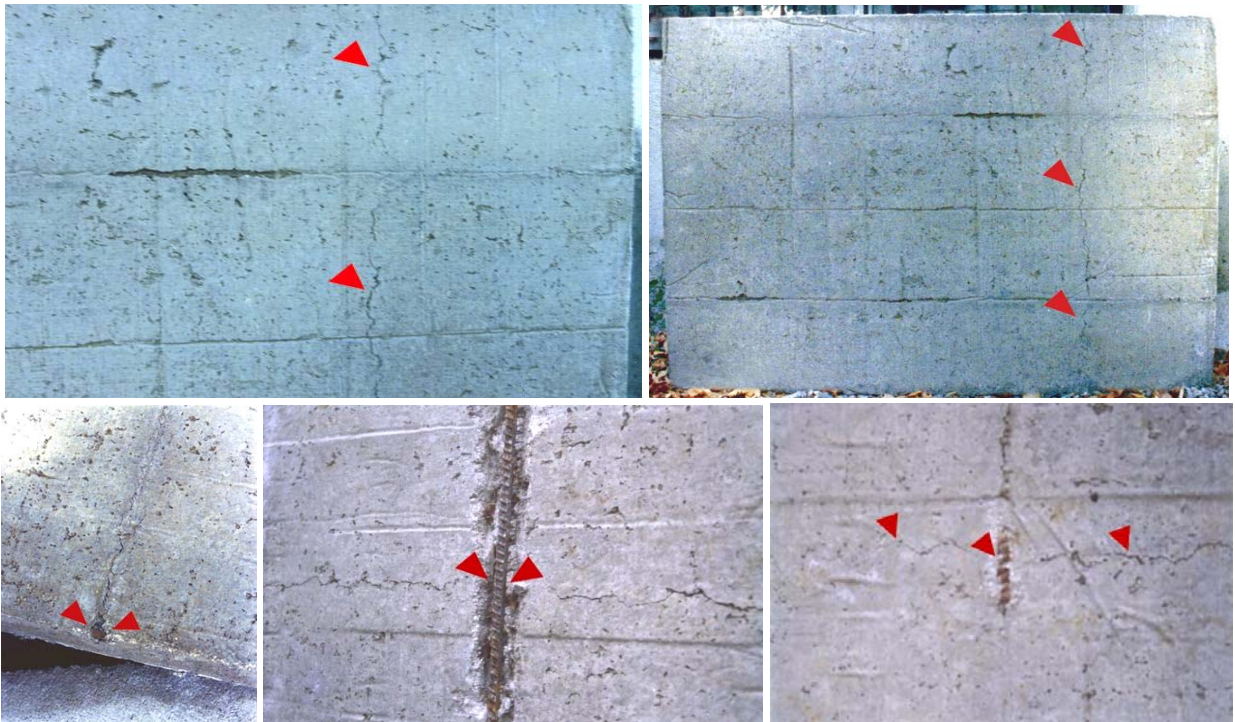

Fig. 2.5. The beneath surface of the tested slab having the supplementary bars after the in-bending loading
- The "extended" crack in the expected area (considering the loading manner) is clear.
- Here, the supplementary bar (with the appropriate involvement in the concrete) has not yielded.

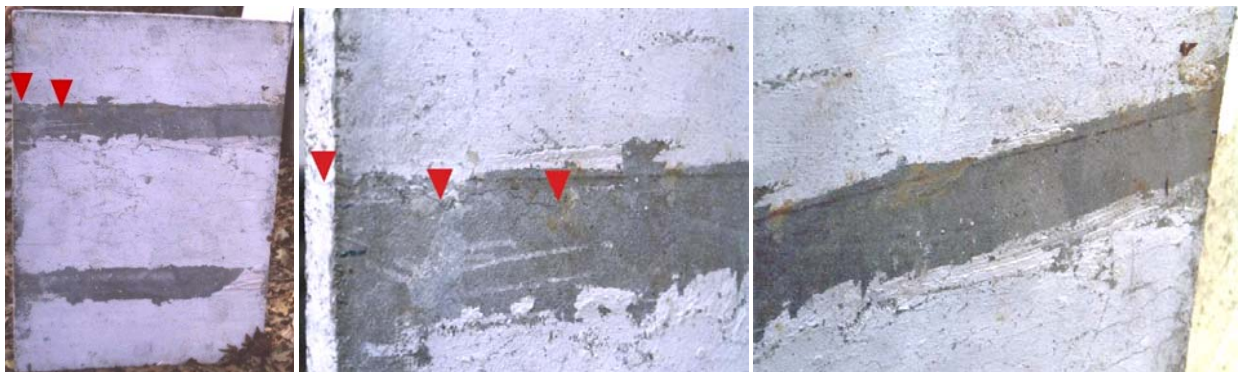

Fig. 2.6. The above surface of the tested slab having the supplementary bars after the in-bending loading
- Only a "limited (non-extended)" crack ("having short length and not deep, and in a certain area under the external edge of the employed girder for loading") is obvious.

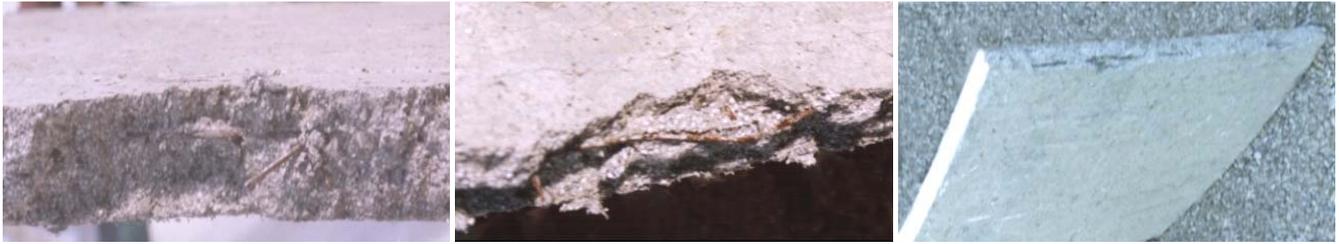

Fig. 2.7. The pieces of the tested slab "not having" any supplementary bar after the in-bending loading

It is worthy of mentioning that; here, in the special lightweight matrix employed for making the ECRLC in these tests (both in the fiber-less and especially, in the used fibered samples of the said lightweight concrete with high strain capability in compression), the stress block indices (α & β) and particularly, the strain correspondent with the final, complete failure ($\varepsilon_{cu}$) in compression were significantly more than those of the usual concretes (so that the behavior of the mentioned special fibered lightweight concrete in compression after the fracture point has been partially similar to the so-called paste-form materials).]

In the performed tests, "even of some primary compressive type in the axial loadings", the fracture pattern has not been brittle. In this regard, occurrence of a type of "being gradually compressed" instead of "the usual outspreading shatter" in the over-threshold in-compressive loadings is considerable. (This is partially similar to the so-called paste-form materials behavior after fracture in the in-compressive loadings.) [Nonetheless, the employment of this certain type of the fibered lightweight concrete (with the said certain lightweight particles and fibers) is not a necessary (inevitable) condition for getting access to the systems generally called as ECRLC. Anyway, while decrease of possibility of the beam fracture of primary compressive type in bending (together with increase of the modulus of resilience and decrease of the density) is a necessary condition to count a compound material as the "Resilient Composite System", non-brittle fracture pattern in "primary compressive fractures" is not a necessary condition in this regard. (However, respect to the mentioned general structural properties and stated requirements for the components of the lightweight fibered matrix used in the RCS, the stress block indices (α & β) and the strain correspondent with the final, complete failure ($\varepsilon_{cu}$) in compression in the fibered matrix employed in these systems could be higher than those of the usual concretes according to the case.)]

Above all, "respect to the mentioned pattern of bending in this system", the fundamental necessity of equivalence of computable compressive and tensile forces resultants during bending ("derived from the usual relations and calculations based upon the basic kinematic assumption of flexure theory") and its resulted trigonometric similarities are unsubstantial. Thereby, the usual calculations of the equilibrium steel amount to attain the low-steel bending sections with non-brittle fracture pattern (of secondary compressive type) and the related usual limitations have no indication of being propounded. As well, the usual strategic restriction in benefiting tensile reinforcements in beams (particularly, as slabs) is eliminated.

As it was pointed before; the usual relations & equations (based upon the basic kinematic flexure theory) employing to calculate the in-bending beam nominal capacity ($M_n$) will result much fewer amounts of the actual amount ($M_e$) in practice. Particularly, the cracking moment ($M_{cr}$) and *"the actual amount of strain in the elastic limit ($\varepsilon_y$)"* and therefore, the modulus of resilience in bending [4], [6] are considerably higher than the amounts calculated by the common equations based upon the basic assumption of flexure theory.

It should be said that; using the supplementary steel bars under the lattices in the tested slabs has been obviously impressive on all parts of the stress-strain diagrams in bending and compressing, such as the slope of the ascending branch (pointed to the "rigidity" and modulus of elasticity), the energy absorption capacity, ductility, strength, fracture toughness, ultimate strength energy [$U_u = (\sigma_y \cdot \sigma_u / 2) \varepsilon_u$] [3], [6], etc.

It is also worthy of mentioning that; even in the compressive **(**column-like) loading of the tested pieces, which steel lattices and additional supplementary steel bars positions in them have been towards convex surface in the axial loading and also regarding the used lightweight concrete's compressive strength, the related "slenderness ratio", and load amount and exertion method, fracture of "primary compressive" type has been occurred during the in-compressive loading of the slabs, yet the fracture pattern of the pieces has been significantly fine (non-brittle) "due to the certain concrete's texture and behavior". (Fig. 3)

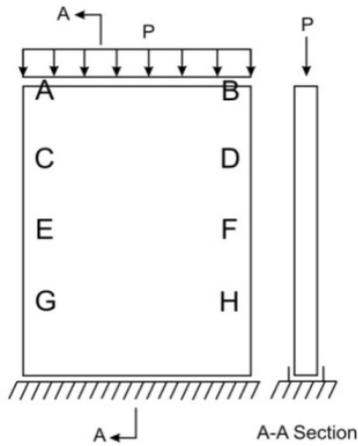

Fig. 3.1. The manner of "in-compressing loading" of the instance of the mentioned slabs (made according to the system as the ECRLC) having the supplementary bars in a manner similar to the ASTM E 72 standard

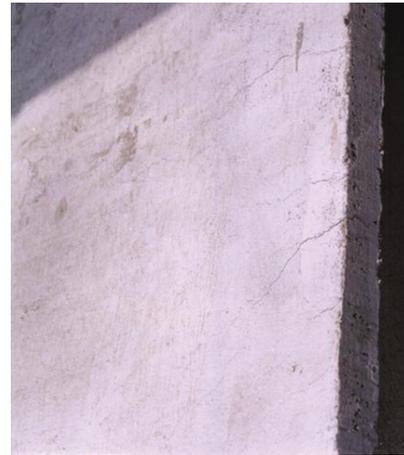

Fig. 3.2. Part of the surface of the tested slab having the supplementary bars after the in-compressive loading - Disseminated compressive cracks are clear.

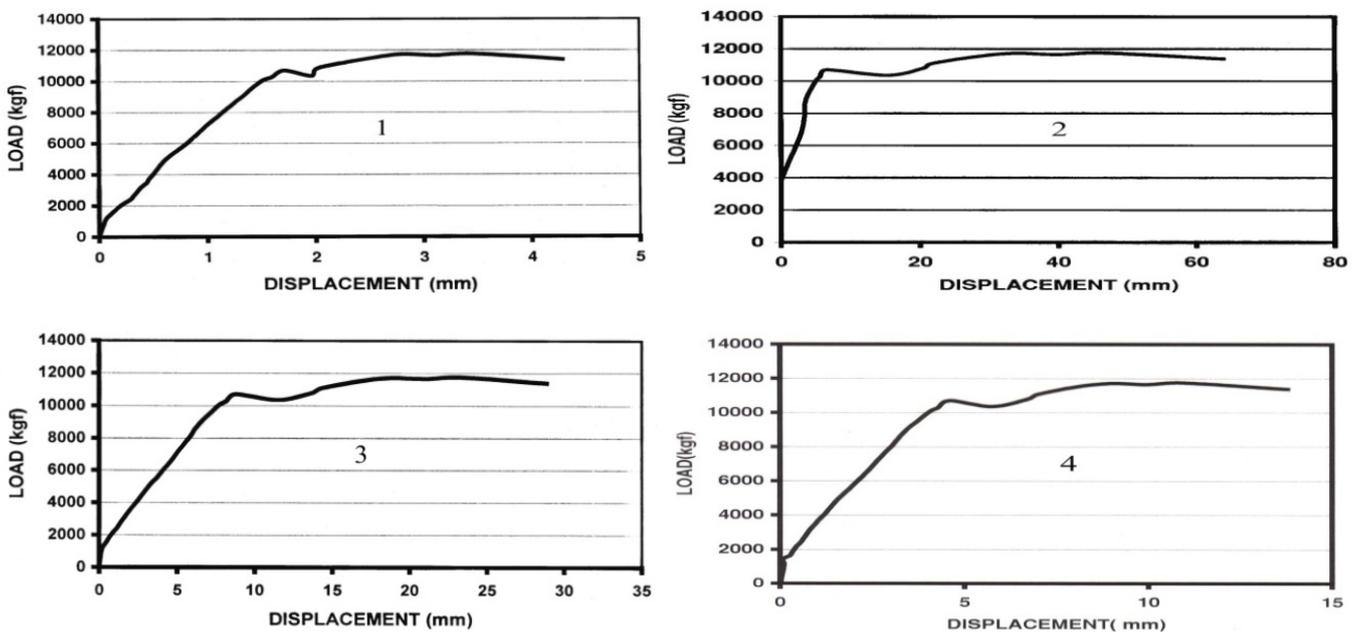

Fig. 3.3. The load-displacement diagrams of the in-compressing loading of an instance of the mentioned slabs (made according to the system as the ECRLC) having the supplementary bars
1- The mean of axial displacement, measured at the points of A & B *(head of the slab)*
2- The mean of lateral displacement, measured at the points of C & D *(3/4)*
3- The mean of lateral displacement, measured at the points of E & F *(1/2; middle of the slab)*
4- The mean of lateral displacement, measured at the points of G & H *(1/4)*
[The stated results have been obtained in spite of the presence of "the unwanted primary, asymmetrical deflection" in the tested slab. The amounts of the mentioned deflections in the "concave" slab before the in-bending and in-compressing loadings, the structure of the system, and the related dimensions have been shown in the afore-presented figures as the figures 2.3 & 2.4. (Naturally, deflections in the slabs increase throughout the loading.)]

In the mentioned systems, which are behaving as an integrative and homogenous beam in their major strains during bending, the elastic strain energy [3] and modulus of resilience in bending considerably increase, and the weight diminishes. As well, regarding the rise of strain energy density, specific capacities of energy reserving and absorption (as the ratio of the capacities to the density) [3], [6] are high. [Generally, noticing to the related issues about the so-called Composite Materials and their particularities, *"Specific* Capacity" is ratio of the capacity to the density.]
. Moreover, the said particulars (such as the special in-bending strain pattern and the role of the "reticular and integrated structure" in: rise of the strength reserving, confronting with formation, development, deepening and changing of the vertical and diagonal cracks, decrease of the accumulating effect of the bending moment and tensile force in the section, and increase of the strength against the piece length alteration) are effective in this system's specific strength increase also in shearing and torsion.

Meanwhile, according to the used components and composition of the said fibered lightweight concrete in the mixture plan (especially, in proportion with each other) and in respect of the texture and qualities of this integrated system and its consistent cement material, the mentioned system could also have suitable "durability" against some of destructive agents, as some ions, in long-term. [7], [8], [9]

## V. SUPPLEMENTARY ELEMENTS

If needed and "according to the case", simultaneously using some auxiliary methods and additional, accompanying elements (such as the supplementary reinforcements, connection strips, foam pieces, reinforcing in different levels, etc) in proportion with this system could be taken into consideration. "However, in general, these supplementary elements are not necessary for counting a system as the so-called Resilient Composite Systems."

For example and as it was mentioned before; employing the supplementary steel bars under the lattices of the tested slabs in the presented experiments has been obviously impressive on all parts of the stress-strain diagrams in bending and compressing, such as ascending branch slope (as the "rigidity" and modulus of elasticity), energy absorption capacity, ductility, strength, fracture toughness, ultimate strength energy, etc.

As another "example", in employing this composite system to make some ceilings as the ceilings having the so-called accessory or secondary beams, the supplementary bars in the slabs (made of this system), in case of having suitable embedment and anchor from two directions (e.g., on the accessory or secondary beams), could improve the construction's totality integration by assisting in combined function of the piece with some other construction elements such as the so-called secondary beams. Meanwhile, fittingly employing these supplementary reinforcements with appropriate anchor and connection advances the under-serving behavior of the beam and could be impressive at primarily fix keeping the suitable condition in the piece (at the time of executing the slab) for better control and accumulation of the stated useful controlled contractile stresses in the reinforced matrix of the system and also in providing: the expedient capability of lateral shearing forces transferring and proportional rigidity in the piece, stronger compressive seats for the said slab on the mentioned secondary beams, and the expedient elastic stability. (Fig. 4)

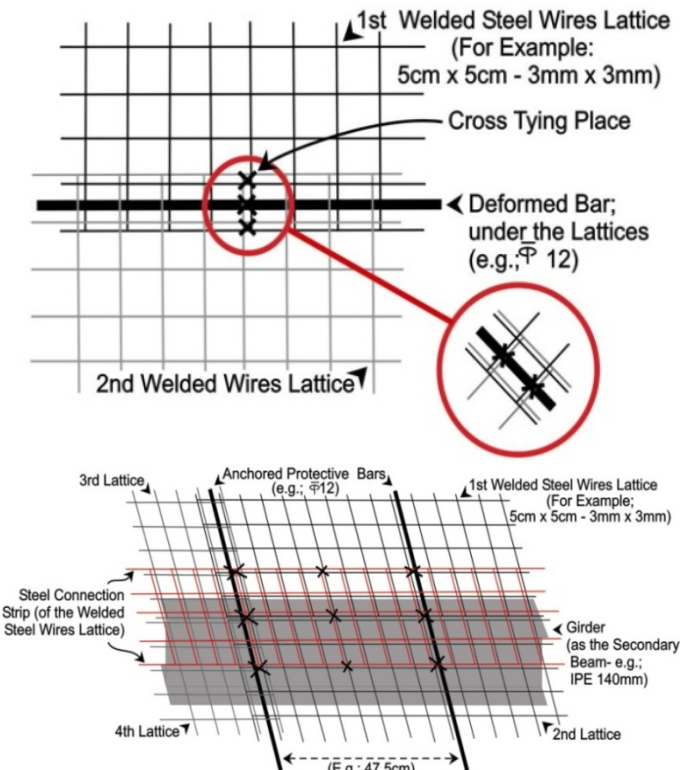

Fig. 4. An Instance of the execution manner of a type of the mentioned ceiling with using the said elements and components (in form of the usual ceilings called as Composite)

Besides to the role of the said composite system's particulars (as highly being of elastic strain capability, modulus of resilience and energy reserving capacity in bending, consistency of the bonding matrix and its suitable involvement with the utilized fibers and other reinforcements in the "integrated, reticular system"), these supplementary elements could, according to the case, be also useful in: partially confronting with immediate and long-term deformities such as loss, creep and fatigue in the piece; increasing fracture toughness, impactive strength and endurance limit (as the ratio of "the fracture strength in frequent load application in the lapse of time" to "the short-term static strength"); increasing the strength in shear and compression; providing expedient elastic stability, expedient lateral shearing forces transferring capability and proportional rigidity in the piece (in the expected cases); more improving the under-serving behavior, etc. ["For example"; in employing this system as the slabs for making the ceilings with the so-called accessory beams, using some strips of the lattices in fitting width, as the connection strips placed at the meeting location of the adjacent meshes of the slabs on the seating place of them (on the accessory beams), could support the applying location of the maximum shearing force in the slab in its turn. (Fig. 4) As well, utilized appropriate foam sheets or blocks underneath the slabs could play the role of a lightweight mold, insulation, and vibration absorbing and intervening according to the case.]

Meanwhile, if needed, it is also possible to expediently employ nanoparticles of active Silica, micro-fibers, Hydrophilic fibers and materials, etc, to more improve the integrity, strength, and duration of the matrix and to more advance the behavior of the structure particularly in bending and against the impacts and continued dynamic loads.

## VI. APPLICATIONS

Considering the subjects and particulars stated for the simple, practical system named as "Elastic Composite, Reinforced Lightweight Concrete" (as a type of the "Resilient Composite Systems" with the stated general structural properties and functional criteria), this system could be efficiently employed as the *"in-bending"* and in-torsion elements and also for construction of the elements performing the acts as absorbing the impacts, shocks, vibrations and dynamic loads (in bending), shielding, etc.

Besides, with respect to the properties as lightness, insulating, durability, work-ability and high forming capability of some components using in this system (such as a special type of lightweight concrete with high strain capability) and regarding the option of employing supplementary elements and auxiliary methods (according to the case), this system and some of its main components, such as the mentioned special lightweight concrete, can be employed in various cases. [7], [8], [9], [10], [11], [12], [13], [14]

These applications are among them:

- Constructing "various types of beams" such as the slabs and lightweight flat, sloped or dome-shaped ceilings (having high resistivity and specific bearing capacity in bending with non-brittle fracture pattern). Thereby, this system can be used in construction of roofs, floors and decks, buildings & towers, multi-floor parking garages, etc.

In addition, it is possible to employ this applied system in construction of bridges, roads, and railroad & subway structures (for instance, as the "Slab Tracks", Traverses and lightweight *"bearing"* or non-bearing vibration and impact absorber elements "under the rails").

Generally, considering the said integrated, reticular system's specialties, as the high energy absorption capacity and appropriate "Endurance Limit" in bending and suitable behavior against the dynamic loads, impacts, and shocks, it could be beneficially utilized to construct various kinds of "the vibration and shock absorber and exposed to continual dynamic loads (in bending) pieces". [In this regard and for instance, employing the Expanded Polystyrene beads with more elasticity and modulus of elasticity (as the HIPS grade) or other appropriate elastomeric beads as the flexible lightweight aggregates in the system could also be taken into consideration according to the case. *Here, it is worthy of mentioning that; regarding the phenomena as fatigue and creep, the mixture plan of the utilized fibered lightweight concrete (as the fibered matrix) could be planned so that its short-term compressive strength be "at least a certain percentage higher than the maximum applied stress" (also considering the "impact coefficient" in dynamic loadings in bending).* Meanwhile, it is clear that; there is a direct, considerable relationship between the modulus of elasticity of the employed lightweight aggregates (within the stated requirements for the components used in the RCS) and the modulus of elasticity of the fibered lightweight concrete. It should be also considered that; the modulus of elasticity of some materials and fibers varies in impact. (For example, the modulus of elasticity of Polypropylene fibers increases through impacts.)];

- Making impact, shock, vibration and expulsion absorbing and intervening pieces and shields, such as lightweight, secure intervening Guards (e.g., the secure side road guards); quiver and bullet protective pieces (with the option of probably employing high-strength fine aggregate concrete or mortar under the mentioned lightweight concrete if needed), etc;

- Producing "lightweight internal and external walls or partitions" (with appropriate steadiness and behavior against the impacts, shocks and blast), which could be also the thermal, humidity and sound insulating or intervening. [7], [8], [9]

For instance, the non-bearing walls, made of the ***"lightweight reinforced sandwich-panels"***, could be constructed by easily executing the mentioned "*non-brittle*", insulating lightweight concrete (as a work-able, paste-form and adhesive material before hardening) on the both sides of the installed foamed steel meshes. The steel meshes having a layer of the foam (between two meshes) could be partially similar to the so-called Tri-dimensional (3-D) or Space Panels having the "fire retarded" Polystyrene sheet as the foam performing the act of insulation (insulator) and mold. (In this application case, we may not need to employ many fibers in the used lightweight concrete according to the case.)

Contrary to some other usual lightweight insulating concretes, this special non-brittle lightweight concrete can be well joined with the meshes to get access to the "non-brittle" lightweight pieces (as the integratedly functioning units) having appropriate steadiness and behavior also against the disseminated (not sharp) impacts and shakes (with capability of absorbing the energy).

**"For example", a proposal mixture plan of the lightweight concrete utilized for this particular simple and practical application could be as follow: Portland cement (type 1 or 2, according to the case) + micro Silica fume (7% of cement materials) = 550kg/m$^3$ - W / C+S = 0.4 - Lignosulfonate powder (as a *low-price* "plasticizer" and retardant) = 1.1kg/m$^3$ - Polypropylene fibers (with Denier 3 and in Length 12mm, which could be blown and mixed**

with "scanty" amount of the micro Silica fume for better separation and distribution of the fibers) = 1.265kg/m$^3$ - Expanded Polystyrene (EPS) beads ($D_{50} \leq 3.2$mm) up to 1m$^3$ (with the lowest "actual density", for example by appropriately increasing time expansion of the suitable fine primary granules). [According to the case, it is also possible to replace the Silica fume with some other *appropriate* "Pozzolanic Materials" [7], [8], [9], [15], [16], [17] accompanied by expediently changing the amounts of the employed Portland cement, Expanded Polystyrene beads, additives, and "W/C Ratio" in the mixture plan.] These materials should be "well" mixed (e.g., for at least 15 minutes). The mentioned fibered lightweight concrete oven-dry density is about 660kg/m$^3$, and the density in the usual normal condition could be about 730kg/m$^3$. This production provides the required particularities also for the in-place use conditions. [Generally, considering the subjects stated in ACI-523.1R & 2R and other related references; for any constructional employment of lightweight concretes (even as the non-bearing elements, not only used to insulate), the lightweight concrete's f´$_c$ should not be less than 2.07mpa, and the drying shrinkage within 90 days in the standard conditions should not be more than 0.2%.] In the afore-said light panels, the welded steel wire meshes and the fire retarded Polystyrene sheet (as insulation and mold) could have variable dimensions according to the case (e.g., 8cm×8cm-3mm×3mm for the steel lattices, and 5cm diameter for the Polystyrene sheet with low density). Regarding the mentioned application as the non-bearing walls in pressure, in this certain case, the usual "distance" between the Polystyrene sheet and the utilized lattices in the panel (which is usually existed in some similar and common panels as the 3D Panels, also employed as the bearing element in compression) is not necessary. As well, in a wide view. These advantages could be among them: enhancement of the bearing capacities; significant reduction in construction weight (sometimes, up to 3-6 times comparing with the case of using some usual heavy materials) and saving in the related expenses; *"improvement of behavior, resistance and safety against earthquake and shake", shock and explosion*; appropriate thermal insulation and sound intervening; increase of indoor useful space (owing to some elements' dimensions reduction); suitable performance and having various possibilities of execution as the in-place, precast and semi-precast implementations (according to the case); etc. (Fig. 5)

The said benefits could have high importance also in constructing high buildings & towers and particularly in constructing in the *seismic* and/or far-reaching areas.

in this case, the connecting steel wires (between the steel lattices placed in two faces of the panel) could be simply perpendicular to the Polystyrene sheet surface. The diameter of the utilized past-form lightweight concrete (having the slump = 0) on each surface of the panel could be about 2cm, and the total weight of the finished non-bearing wall (constructed with these lightweight and insulating, "non-brittle", reinforced sandwich panels) with some expedient thin covers (plasters) could be about 50kg/m$^2$. (More information in this regard has been also presented in the related literature. [7], [8], [9]) Considering an existent partially similar and common method of constructing as employing the combination of tri-dimensional (3-D) foamed steel lattices with the usual sand-cement mortar, this practical implementation method could easily spread.]

Employing these types of walls could have some advantages such as: providing more rapidity & easiness in transportation and installation; little materials wasting in implementation; least required additional plasters (noticing the suitable surface of the executed mentioned lightweight concrete); appropriate work-abilities as possibilities of cutting, nailing, rasping (also after hardening of the cement paste, and in order to reach to the leveled surfaces if needed), holding screws, roll-plaques & roll-bolts, and having the capabilities of repair, installation transferring, establishment of frames, doors & windows, execution of various coatings & paints, and adaptation with diverse architectural designs (e.g., in curve surfaces and forms), and expedient flexibility.

Besides, if needed, it is also possible to produce the dry materials (Dry Mix) in standard packs to finally make an ameliorated, workable, fibered lightweight cemental paste for simple execution on the foamed steel meshes.

Generally, there could be several benefits in using this simple and practical system and its components in constructing buildings especially

Here, it is also worthy of mentioning that; considering the properties of this integrated, lightweight structure, the mentioned system or some of its components could be beneficially employed for *Lightweight and Integrated Construction* in the *"seismic areas"* (even sometimes with employing the common used equations based upon the assumption of "linearly being of the strain changes in the beam height during bending"). Thus, "*simple and handy application of the mentioned structures and related components and materials to effectively increase the resistance and safety against earthquake*" could be taken into consideration particularly in the seismic areas. Lightness, high resilience and capacities of energy absorption and reserving in bending, secure fracture pattern, appropriate behavior against the disseminated high impacts and shakes, suitable integrity, and not utilizing of high weight and separated materials with discordant behavior are among the specifics, which are important in this regard. *[As a general rule; in many cases, "Lightweight and Integrated Construction" could be counted as the main, "practical" approach to effectively increasing the resistance and safety of constructions against earthquake and lateral forces, in large extent.]* [7], [8], [9], [13], [14]

 - Constructing some energy absorbing bases (for example, in the integrated or sandwich forms);
 - Using in marine structures and floaters;
 - Employing in making the structures and objects such as; lightweight pipes and ducts in various forms (with appropriate capability of energy and vibration absorption); some intervening & separating pieces and partitions

 (such as the guards against spilling); springhouses, containers, and reservoirs; some lightweight pieces, covers, volumes, and objects in various shapes, such as the lightweight facade pieces, lumbers, cabinet, counter, etc; lightweight vibration & sound intervening & isolating partitions and pieces (such as the intervening walls against persecuting noises in the electricity transmission sites);
 - Etc. [Some of the other potential application cases of the mentioned system and its components (as the said special lightweight concrete with possibility of utilizing various kinds of lightweight aggregates & particles, fibers and cement materials) are: employing in production of: non-bearing fire protective and intervening pieces and covers (with appropriately attention to the expedient requirements in the used components); lightweight suspended ceilings; other kinds of sandwich panels; lightweight flooring, plastering and insulating; some cases of filling and sloping, clogging some crossing points and crevices having movement….]

In the Fig. 5, some application cases of the pointed materials and the system's components in some usual methods have been shown. (Fig. 5)

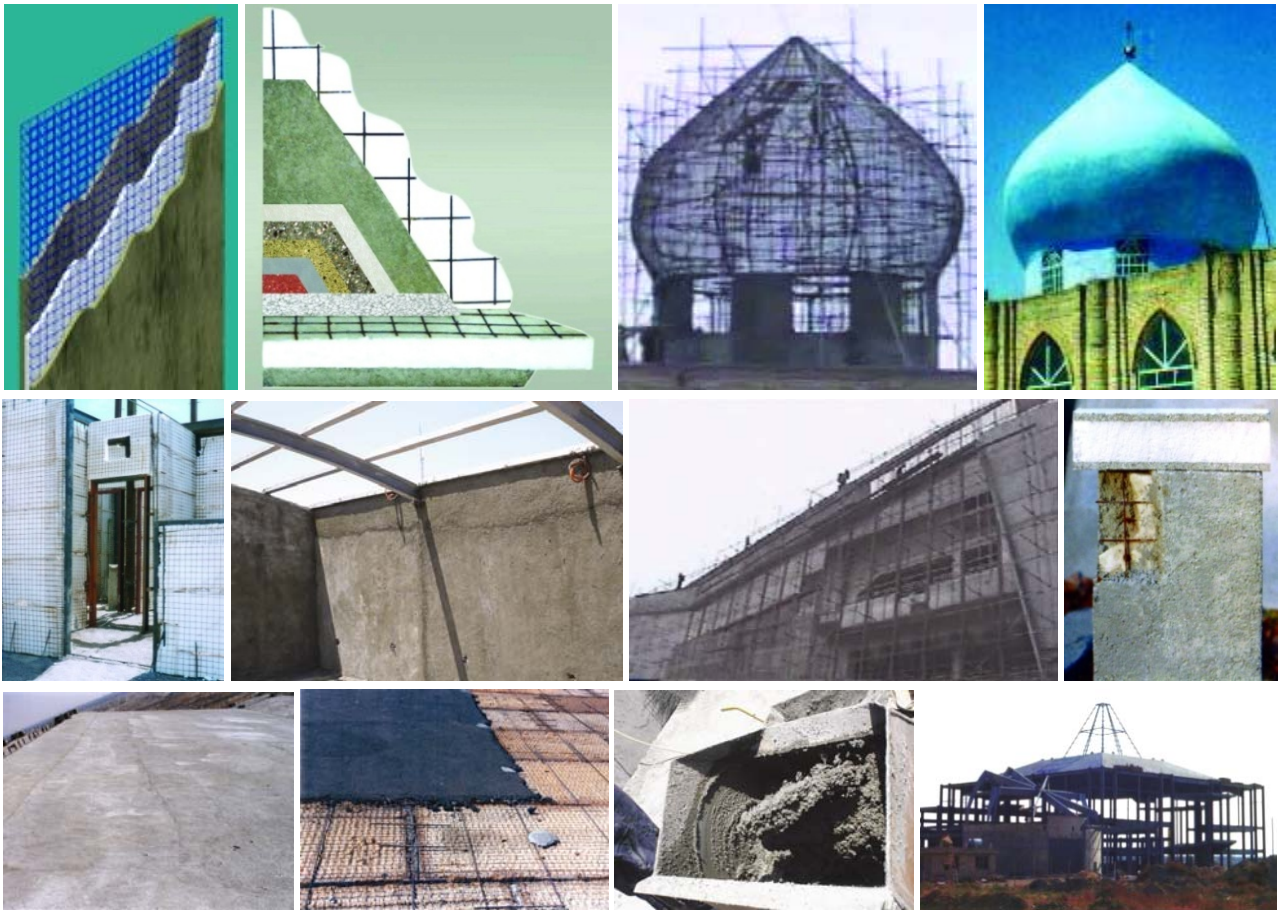

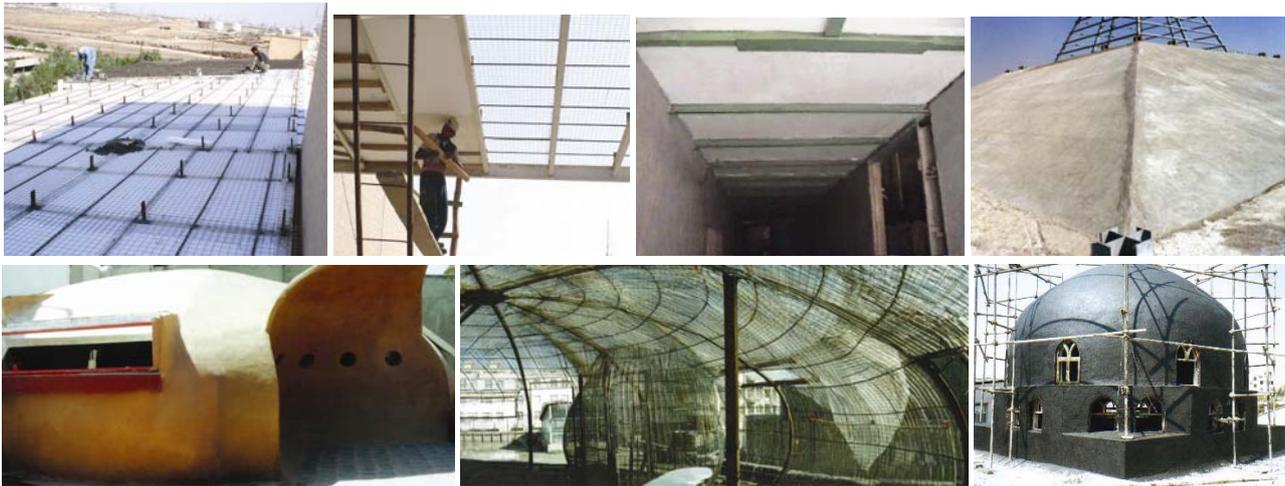

Fig. 4. Some application cases of the pointed materials and the system's components in some usual methods

## VII. FINAL REVIEW

A kind of "Elastic Composite, Reinforced Lightweight Concrete" with the said specifics is a type of the "Resilient Composite Systems (R.C.S.)" in which, contrary to the basic geometrical assumption of flexure theory in the Solid Mechanics, the strain changes in the beam height during bending is typically "Non-linear".

Indeed, the RCS, as the Elastic Composite, Reinforced Lightweight Concrete (ECRLC), do not behave as most of the solid materials in bending.

In the "Resilient Composite Systems", distributed pores and/or appropriate lightweight aggregates or beads, accompanied by the reticular structure of the strengthened conjoined matrix, bring about the expedient internal shape changes during bending and continuing the elasticity in bending with the said nonlinearly pattern. This means better distribution of the stresses and strains and better utilizing the potential capacities of the employed reinforcements in bending and tension; whereas, in the usual lightweight concretes for instance, distributed hollow pores (such as the gas bulbs in the cellular concretes) or lightweight aggregates (such as Plastic, Rubber or polystyrene beads or any other kind of lightweight aggregates such as Perlite and Vermiculite) decrease the modulus of resilience in bending and could increase "the possibility of beam fracture of brittle and primary compressive type" in bending (compared to the concrete with higher density) according to the case.

In this way, by using the mentioned method to make the said particular composite systems, we could considerably increase the modulus of resilience and bearing capacity in bending "together with" significant decrease of the weight and also possibility of beam fracture of primary compressive type. Through making these particular integrated functioning systems, for the first time, the said (paradoxical) properties have been concomitantly fulfilled in "one functioning unit" altogether.

Respect to the special pattern of the strain changes during bending in the particular Resilient Composite System called as the ECRLC, this system as an integrated functioning unit with the reticular arrangement and texture has more strain capability (particularly within the elastic limit), energy absorption and load bearing capacities in bending compared to the usual reinforced concrete beams.

Thereby, through employing this applied structure, solving some of the main problems in lightweight concretes application, especially the deadlock of brittle and insecure being of fracture pattern in many of the usual reinforced lightweight concrete structures, is provided; reaching to the high bearing capacities in bending elements (even with low dimensions & weights) is to hand, and getting access to a simple and practical opportunity for "qualitative development of capabilities of using lightweight concretes" (especially with oven-dry densities of < 1350-1400kg/$m^3$ and compressive strengths of <14-17mpa, and even with oven-dry densities of < 800kg/$m^3$) is conceivable.

Naturally, by more studies in the field, these structures and their applications in various fields could be developed more.

ACKNOWLEDGEMENTS

Author wishes to thank all people, who helped in reaching the presented findings and subjects in various ways; as Dr. H. Rahimi (Prof. of Irrigation and Reclamation Engineering Department of Tehran University Agricultural College), Mrs. M. Esmaeili (Civil Engineer), Mrs. R. Bahramlueian, and the other ones, for their valuable comments and assistances.